\documentclass[sigconf, authorversion, nonacm]{acmart}

\AtBeginDocument{%
  }

\setcopyright{acmlicensed}
\copyrightyear{2026}
\acmYear{2026}
\acmDOI{XXXXXXX.XXXXXXX}
\acmConference[AGENT '26]{International Workshop on Agentic Engineering}{April 12--18, 2026}{Rio de Janeiro, Brazil}
\acmISBN{978-1-4503-XXXX-X/26/04}

\begin{document}

\title{Does SWE-Bench-Verified Test Agent Ability or Model Memory?}


\author{Thanosan Prathifkumar}
\affiliation{%
  \institution{Central Peel Secondary School}
  \city{Brampton}
  \country{Canada}}
\email{p.thanosan23@gmail.com}

\author{Noble Saji Mathews}
\affiliation{%
  \institution{University of Waterloo}
  \city{Waterloo}
  \country{Canada}}
\email{noblesaji.mathews@uwaterloo.ca}

\author{Meiyappan Nagappan}
\affiliation{%
  \institution{University of Waterloo}
  \city{Waterloo}
  \country{Canada}}
\email{mei.nagappan@uwaterloo.ca}

\renewcommand{\shortauthors}{Prathifkumar et al.}

\begin{abstract}
  SWE-Bench-Verified, a dataset comprising 500 issues, serves as a de facto benchmark for evaluating various large language models (LLMs) on their ability to resolve GitHub issues. But this benchmark may overlap with model training data. If that is true, scores may reflect training recall, not issue-solving skill. To study this, we test two Claude models that frequently appear in top-performing agents submitted to the benchmark. We ask them to find relevant files using only issue text, and then issue text plus file paths. We then run the same setup on BeetleBox and SWE-rebench. Despite both benchmarks involving popular open-source Python projects, models performed 3 times better on SWE-Bench-Verified. They were also 6 times better at finding edited files, without any additional context about the projects themselves. This gap suggests the models may have seen many SWE-Bench-Verified tasks during training. As a result, scores on this benchmark may not reflect an agent’s ability to handle real software issues, yet it continues to be used in ways that can misrepresent progress and lead to choices that favor agents that use certain models than strong agent design. Our setup tests the localization step with minimal context to the extent the task should be logically impossible to solve. Our results show the risk of relying on older popular benchmarks and support the shift toward newer datasets built with contamination in mind.
  
\end{abstract}



\maketitle

\section{Introduction}
Large language models (LLMs) have greatly reshaped software engineering. To measure the capabilities of LLMs and to compare different LLMs, researchers have created benchmarks to standardize the process. One popular task that LLMs can be used for is fixing issues within a codebase. SWE-Bench-Verified has become the de facto benchmark to measure and compare the performance of LLM-based systems in fixing issues \cite{jimenez_swe-bench_2024}. 
The corresponding leaderboard ranks these systems based on the percentage of issues they can resolve.

SWE-Bench-Verified is entirely built on open-source GitHub repositories. However, this can be a problem. This data is likely to have been seen by LLMs, as many of these LLMs (such as GPT-4 and Claude) are trained on large corpora of the internet \cite{aleithan_swe-bench_2024, zhou_lessleak-bench_2025}. This would make the leaderboard and resolution rates shown unreliable, as the underlying model may memorize examples rather than learn generalizable reasoning patterns. In other words, the model may appear intelligent due to memorization rather than genuine learning.
This motivates us to analyze the extent to which SWE-Bench-Verified is leaked into LLMs. In this paper, we assess whether the benchmark itself, SWE-Bench-Verified, has been leaked into one of the top-ranking LLMs, Claude Sonnet. 

To better understand the extent of leakage, we conduct an experiment. In our experiment, we aim to solve the localization task that precedes bug fixing. The LLM needs to know where to fix the bug before creating a patch that targets the identified locations. We intentionally handicap the model by giving it insufficient input. We then compare the performance on SWE-Bench-Verified, BeetleBox \cite{chakraborty_blaze_2025}, and SWE-rebench \cite{badertdinov_SWE-rebench_2025}. BeetleBox is a large cross-language dataset that consists of numerous repositories. We select 5 Python repositories that are not in SWE-Bench-Verified at random. Additionally, SWE-rebench consists of continuously updated versions of the SWE-Bench benchmark. We use the SWE-rebench split from January 2025 and September 2025. All three datasets are derived from popular open-source projects. We investigate the localization performance of the top-performing models across these datasets. Several-fold stronger performance with minimal context on SWE-Bench-Verified than on similar datasets like BeetleBox and SWE-rebench suggests the tasks may have leaked into the training data.




\section{Background and Related Work}


\subsection{Data Leakage in SWE-Bench}
Aleithan et al. \cite{aleithan_swe-bench_2024} show that 94\% of the instances in SWE-Bench consist of issues that were created before the training cut-offs of the most common LLMs. This means that the issues from SWE-Bench have probably been seen before by the LLM, as these issues would be in their pretraining data. To test this hypothesis, Aleithan et al. developed SWE-Bench+, a coding benchmark that uses GitHub issues that were created after the training cutoff dates for popular LLMs. On SWE-Bench+, the resolution rates dropped significantly in comparison to SWE-Bench, showing that there is clear data leakage in popular LLMs.

Zhou et al. \cite{zhou_lessleak-bench_2025} show that SWE-Bench has a high leakage rate to open-source models (specifically, StarCoder), compared to other Python benchmarks. In the case of SWE-Bench, Zhou et al. state that the reason lies in the fact that SWE-Bench is collected from GitHub issues, which are included in StarCoder’s pre-training data. This results in an overlap. To better quantify the leakage rate to StarCoder, they have shown that there is a 10.6\% data leakage in SWE-Bench Verified and an 8.7\% data leakage in SWE-Bench. They do this by identifying duplicate data between the pre-training data and SWE-Bench automatically and checking them manually with experts. They believe that the main reasons for data leakage to StarCoder are due to benchmark data being used in pre-training datasets, the reliance on specific platforms (e.g. LeetCode), common data sources (such as GitHub), and overlap between numerous source repositories. 

Liang et al. \cite{liang_swe-bench_2025} test whether LLMs actually reason through issues or they memorize the SWE-Bench tasks. To do this, they make models attempt to identify which files are buggy through only the GitHub issue descriptions. Any information about the file structure is not given in the prompt. They also conduct a cross-repository analysis to compare the performance on SWE-Bench-Verified and other external repositories. In their results, they show that OpenAI’s o3 model achieves a 76\% accuracy on SWE-Bench without any contextual information. This suggests potential data contamination. Models consistently achieved high accuracy of SWE-Bench-Verified with an accuracy of 60-76\%, suggesting that the models are memorizing. When comparing SWE-Bench performance to external benchmarks without any of the repositories from SWE-Bench, accuracy is consistently below 53\%. Additionally, when tested on repositories consisting of other languages, such as C\#, the performance decreases by up to 47 percentage points.

Unlike past studies, which compared the performance of models on SWE-Bench-Verified and other benchmarks on fix generation given the issue and the repo, our study changes the input settings and focuses on the localization phase in a minimal context setting. 
Due to this change in the settings, we can infer a stronger claim of the tasks from the benchmark itself being leaked rather than just the benchmark using leaked data by virtue of being derived from popular open-source projects.

\subsection{Other Faults in SWE-Bench}
Aleithan et al. \cite{aleithan_swe-bench_2024} also show that weak test cases in SWE-Bench may contribute to abnormally high resolution rates. Wang et al. \cite{wang_are_2025} state that SWE-Bench only uses tests that were modified in the pull-request. When executing all the tests instead of only the tests modified in the pull request, there is a 4.5\% performance drop. In fact, 28.6\% samples are obviously incorrect, but yet, they still pass the test suites, showing that these tests are weak. Furthermore, Yang et al. \cite{yang_swe-bench_2024}, Zan et al. \cite{zan_swe-bench-java_2024}, and Zan et al. \cite{zan_multi-swe-bench_2025} provide an additional critique on SWE Bench. These aforementioned papers state that SWE-Bench is exclusively Python, forcing models to overfit on Python-based projects. This prevents other use cases from being represented. In fact, Yang et al. show that top performing models that did well on SWE-Bench tend to struggle with SWE-Bench Multimodal (which are tasks from JavaScript repositories). This gap prevents the benchmark from truly representing complex software engineering tasks in the real world. In contrast, our work is to examine if there is data leakage and not necessarily if the benchmark itself is of good quality. 

\section{Methodology}





\subsection{Models}
We chose two models from the Claude Sonnet series - 3.5 and 3.7. These are state-of-the-art models used by several top-ranking agents in the SWE-Bench-Verified leaderboard. While Claude 4 was released in May 2025, it uses a more recent knowledge cutoff (March 2025, as compared to November 2024). Hence, we believe it should be sufficient to show that data has been leaked into the previous version of the model.

\subsection{Data}
\begin{itemize}
    \item We use all 500 issues from SWE-Bench Verified. 
    \item We randomly select 5 Python-based repositories from BeetleBox. These repositories are the following: Ansible, Apache Airflow, PostHog, Localstack, and Langchain. Next, we select 100 random issues from each of the chosen repositories from BeetleBox (totalling up to 500 issues from BeetleBox). To make sure that these issues are of high quality, each of the issues are manually processed to ensure that they are not vague. The criterion by which this was evaluated was whether the issue had sufficient detail for another developer to look at the problem, reproduce the problem, and take steps to resolve it without the need for further questions. For the issues that were considered vague, we dropped them, and random issues were sampled again until there is a total of 500 issues from BeetleBox that are considered not vague. This manual analysis was carried out by the first author of this paper.
    \item In order to improve the generalizability of our analysis even further, we select the January 2025 split from SWE-rebench and the September 2025 split from SWE-rebench \cite{badertdinov_SWE-rebench_2025}. The January 2025 split consists of 109 issues. The September 2025 split consists of 50 issues.
\end{itemize}

\subsection{Input Settings}
We carry out our experiments with two different input settings. We ask the model for the files that are most likely to have the fix for the issue, i.e., localization. Both these settings have insufficient data to rationally reason about and solve the localization task let alone the task of fixing bugs. 

\begin{itemize}
    \item Issue + File Structure: In this setting, we give as input to the model the issue and the file structure of the repository that the issue belongs to. We do not give the model access to the file itself. The model only knows the names of the files in the repository and the file paths. 
    \item Issue only: In this setting, we provide the models with just the issue and no file names or structure. 
\end{itemize}


\subsection{Evaluation}
We then evaluate the output of the models, which are the file names where the issue will likely be fixed. We do this by comparing against the ground truth in the benchmarks. We carry out two kinds of evalautions: (1) what percentage of issues have all the files in the ground truth present in the model output and (2) what percentage of issues have at least one file from the ground truth in the model output.



\section{Results}

In Table \ref{tab:claude_all_ticket_filename} and in Table \ref{tab:claude_some_filename}, we can see the results when both the ticket and the file structure are given as input. Both models are approximately 4 times better in predicting all ground truth files in the SWE-Bench-Verified benchmark when compared to BeetleBox, and approximately 2 times better when compared to SWE-rebench. In BeetleBox and SWE-rebench, the poor results are comparable to when the SWE-Bench-Verified benchmark was first released. When predicting at least one ground truth file, we can see a similar higher performance in SWE-Bench-Verified, albeit to a smaller extent when compared to BeetleBox and SWE-rebench.

\begin{table}[ht]
  \caption{Percentage of issues with all ground truth files in the output when given Ticket + File Structure}
  \label{tab:claude_all_ticket_filename}
  \begin{tabular}{lcc}
    \toprule
    Model & Claude 3.5 & Claude 3.7 \\
    \midrule
    SWE-Bench & 76\% & 73\% \\
    BeetleBox & 21\% & 17.6\% \\
    SWE-rebench (09/2025) & 28\% & 22\% \\
    SWE-rebench (01/2025) & 43.12\% & 44.04\% \\
    \bottomrule
  \end{tabular}
\end{table}

\begin{table}[ht]
  \caption{Percentage of issues with at least one ground truth file in the output when given Ticket + File Structure}
  \label{tab:claude_some_filename}
  \begin{tabular}{lcc}
    \toprule
    Model & Claude 3.5 & Claude 3.7 \\
    \midrule
    SWE-Bench & 83.60\% & 82.40\% \\
    BeetleBox & 68.6\% & 59.8\% \\
    SWE-rebench (09/2025) & 74\% & 70\% \\
    SWE-rebench (01/2025) & 87.16\% & 81.65\% \\
    \bottomrule
  \end{tabular}
\end{table}


\begin{table}[ht]
  \caption{Percentage of issues with all ground truth files in the output when given Ticket only}
  \label{tab:claude_all_ticket}
  \begin{tabular}{lcc}
    \toprule
    Model & Claude 3.5 & Claude 3.7 \\
    \midrule
    SWE-Bench & 65\% & 63.20\% \\
    BeetleBox & 12.2\% & 12\% \\
    SWE-rebench (09/2025) & 12\% & 8\% \\
    SWE-rebench (01/2025) & 17.43\% & 19.27\% \\
    \bottomrule
  \end{tabular}
\end{table}

\begin{table}[ht]
  \caption{Percentage of issues with at least one ground truth file in the output when given Ticket only}
  \label{tab:claude_some_ticket}
  \begin{tabular}{lcc}
    \toprule
    Model & Claude 3.5 & Claude 3.7 \\
    \midrule
    SWE-Bench & 72.80\% & 71\% \\
    BeetleBox & 43\% & 44.6\% \\
    SWE-rebench (09/2025) & 34\% & 30\% \\
    SWE-rebench (01/2025) & 34.86\% & 38.53\% \\
    \bottomrule
  \end{tabular}
\end{table}

In Table \ref{tab:claude_all_ticket} and Table \ref{tab:claude_some_ticket}, 
we present the result under the input setting with only the issue information. When evaluating against all ground truth files, we can see that the performance in SWE-Bench-Verified is almost 6 times that of BeetleBox.  Compared to SWE-rebench, performance on SWE-Bench is approximately 3 times better. Similarly, when considering at least one file from the ground truth, the performance is still higher in SWE-Bench-Verified than in BeetleBox and SWE-rebench. We also notice that the difference when considering the issue only is much larger than when considering the issue + file structure. 

We can make a few observations from taking the results in all 4 tables:
\begin{itemize}
    \item Results in SWE-Bench-Verified are always significantly better despite both sets of issues being derived from popular Python projects.
    \item The difference is more when considering all ground truth files than just one file. We hypothesize that the model may be able to determine some files by chance, mention in the ticket, semantics in the file name or by virtue of being open-source projects. But when all files need to be correctly localized, SWE-Bench-Verified is better by several folds. This ability to find all the files only in a popular benchmark, even though the data in the benchmark is from popular open-source projects, is hard to justify.
    \item The raw performance in each comparative setting is lower when considering issues alone, when compared to issue + file structure. This is expected as there is much less context when considering the issue alone. If LLMs have internalized the contents and file structure in open-source repositories, this should translate to other popular projects as well. 
    \item However, the difference between SWE-Bench-Verified and BeetleBox is even larger when only the issue is considered. This is the setting with the greatest handicap. Of course, we can expect the performance to be bad. And we do see that the performance is bad in the BeetleBox benchmark. But the performance in the SWE-Bench-Verified benchmark is exceedingly high. 
\end{itemize}

The contrast between the high performance on SWE-Bench-Verified and the weak results on other similar datasets points to the chance that the models have seen these tasks before. This may reflect memorization rather than real problem-solving.


\section{Conclusion}

We find that both Claude models perform very well on localization in SWE-Bench-Verified even when given only the issue text, with no file names or code. In contrast, performance drops sharply on BeetleBox and SWE-rebench, even though they also use popular open-source Python projects. This gap suggests recall of benchmark tasks rather than general skill. If the benchmark were testing real learning, we would expect stronger transfer across similar projects, which we do not see.
These results indicate that SWE-Bench-Verified data is likely present in model training. As a result, scores on this benchmark may not reflect true agent ability on real software issues. Continued reliance on it risks overstating progress and may lead teams to prefer systems that benefit from training overlap rather than strong agent design.
Benchmarks that refresh tasks and limit contamination, such as SWE-rebench \cite{badertdinov_SWE-rebench_2025}, offer a more reliable path for evaluating software agents. Our findings support a shift toward such settings and highlight the need for care when interpreting results from fixed, widely-used datasets.




\section{Data Availability}
Data, logs and scripts from our experiments can be found at: \url{https://osf.io/q5y47/?view_only=7785a3f32db841d4bd28d680c2c1771c}

\bibliographystyle{ACM-Reference-Format}
\bibliography{paper}

@misc{aleithan_swe-bench_2024,
	title = {{SWE}-{Bench}+: {Enhanced} {Coding} {Benchmark} for {LLMs}},
	shorttitle = {{SWE}-{Bench}+},
	url = {http://arxiv.org/abs/2410.06992},
	doi = {10.48550/arXiv.2410.06992},
	urldate = {2025-05-31},
	publisher = {arXiv},
	author = {Aleithan, Reem and Xue, Haoran and Mohajer, Mohammad Mahdi and Nnorom, Elijah and Uddin, Gias and Wang, Song},
	month = oct,
	year = {2024},
	note = {arXiv:2410.06992 [cs]},
}

@misc{zhou_lessleak-bench_2025,
	title = {{LessLeak}-{Bench}: {A} {First} {Investigation} of {Data} {Leakage} in {LLMs} {Across} 83 {Software} {Engineering} {Benchmarks}},
	shorttitle = {{LessLeak}-{Bench}},
	url = {http://arxiv.org/abs/2502.06215},
	doi = {10.48550/arXiv.2502.06215},
	urldate = {2025-05-31},
	publisher = {arXiv},
	author = {Zhou, Xin and Weyssow, Martin and Widyasari, Ratnadira and Zhang, Ting and He, Junda and Lyu, Yunbo and Chang, Jianming and Zhang, Beiqi and Huang, Dan and Lo, David},
	month = feb,
	year = {2025},
	note = {arXiv:2502.06215 [cs]},
}

@misc{zan_swe-bench-java_2024,
	title = {{SWE}-bench-java: {A} {GitHub} {Issue} {Resolving} {Benchmark} for {Java}},
	shorttitle = {{SWE}-bench-java},
	url = {http://arxiv.org/abs/2408.14354},
	doi = {10.48550/arXiv.2408.14354},
	urldate = {2025-05-31},
	publisher = {arXiv},
	author = {Zan, Daoguang and Huang, Zhirong and Yu, Ailun and Lin, Shaoxin and Shi, Yifan and Liu, Wei and Chen, Dong and Qi, Zongshuai and Yu, Hao and Yu, Lei and Ran, Dezhi and Zeng, Muhan and Shen, Bo and Bian, Pan and Liang, Guangtai and Guan, Bei and Huang, Pengjie and Xie, Tao and Wang, Yongji and Wang, Qianxiang},
	month = aug,
	year = {2024},
	note = {arXiv:2408.14354 [cs]},
}

@misc{yang_swe-bench_2024,
	title = {{SWE}-bench {Multimodal}: {Do} {AI} {Systems} {Generalize} to {Visual} {Software} {Domains}?},
	shorttitle = {{SWE}-bench {Multimodal}},
	url = {http://arxiv.org/abs/2410.03859},
	doi = {10.48550/arXiv.2410.03859},
	urldate = {2025-05-31},
	publisher = {arXiv},
	author = {Yang, John and Jimenez, Carlos E. and Zhang, Alex L. and Lieret, Kilian and Yang, Joyce and Wu, Xindi and Press, Ori and Muennighoff, Niklas and Synnaeve, Gabriel and Narasimhan, Karthik R. and Yang, Diyi and Wang, Sida I. and Press, Ofir},
	month = oct,
	year = {2024},
	note = {arXiv:2410.03859 [cs]},
}

@misc{zan_multi-swe-bench_2025,
	title = {Multi-{SWE}-bench: {A} {Multilingual} {Benchmark} for {Issue} {Resolving}},
	shorttitle = {Multi-{SWE}-bench},
	url = {http://arxiv.org/abs/2504.02605},
	doi = {10.48550/arXiv.2504.02605},
	urldate = {2025-05-31},
	publisher = {arXiv},
	author = {Zan, Daoguang and Huang, Zhirong and Liu, Wei and Chen, Hanwu and Zhang, Linhao and Xin, Shulin and Chen, Lu and Liu, Qi and Zhong, Xiaojian and Li, Aoyan and Liu, Siyao and Xiao, Yongsheng and Chen, Liangqiang and Zhang, Yuyu and Su, Jing and Liu, Tianyu and Long, Rui and Shen, Kai and Xiang, Liang},
	month = apr,
	year = {2025},
	note = {arXiv:2504.02605 [cs]},
}

@misc{liang_swe-bench_2025,
	title = {The {SWE}-{Bench} {Illusion}: {When} {State}-of-the-{Art} {LLMs} {Remember} {Instead} of {Reason}},
	shorttitle = {The {SWE}-{Bench} {Illusion}},
	url = {http://arxiv.org/abs/2506.12286},
	doi = {10.48550/arXiv.2506.12286},
	urldate = {2025-07-01},
	publisher = {arXiv},
	author = {Liang, Shanchao and Garg, Spandan and Moghaddam, Roshanak Zilouchian},
	month = jun,
	year = {2025},
	note = {arXiv:2506.12286 [cs]},
}

@misc{wang_are_2025,
	title = {Are "{Solved} {Issues}" in {SWE}-bench {Really} {Solved} {Correctly}? {An} {Empirical} {Study}},
	shorttitle = {Are "{Solved} {Issues}" in {SWE}-bench {Really} {Solved} {Correctly}?},
	url = {http://arxiv.org/abs/2503.15223},
	doi = {10.48550/arXiv.2503.15223},
	urldate = {2025-07-01},
	publisher = {arXiv},
	author = {Wang, You and Pradel, Michael and Liu, Zhongxin},
	month = mar,
	year = {2025},
	note = {arXiv:2503.15223 [cs]
version: 1},
}

@article{chakraborty_blaze_2025,
	title = {{BLAZE}: {Cross}-{Language} and {Cross}-{Project} {Bug} {Localization} via {Dynamic} {Chunking} and {Hard} {Example} {Learning}},
	issn = {0098-5589, 1939-3520, 2326-3881},
	shorttitle = {{BLAZE}},
	url = {http://arxiv.org/abs/2407.17631},
	doi = {10.1109/TSE.2025.3579574},
	urldate = {2025-07-02},
	journal = {IEEE Transactions on Software Engineering},
	author = {Chakraborty, Partha and Alfadel, Mahmoud and Nagappan, Meiyappan},
	year = {2025},
	note = {arXiv:2407.17631 [cs]},
	pages = {1--14},
}

@misc{jimenez_swe-bench_2024,
	title = {{SWE}-bench: {Can} {Language} {Models} {Resolve} {Real}-{World} {GitHub} {Issues}?},
	shorttitle = {{SWE}-bench},
	url = {http://arxiv.org/abs/2310.06770},
	doi = {10.48550/arXiv.2310.06770},
	urldate = {2025-07-03},
	publisher = {arXiv},
	author = {Jimenez, Carlos E. and Yang, John and Wettig, Alexander and Yao, Shunyu and Pei, Kexin and Press, Ofir and Narasimhan, Karthik},
	month = nov,
	year = {2024},
	note = {arXiv:2310.06770 [cs]},
}

@misc{badertdinov_swe-rebench_2025,
	title = {{SWE}-rebench: {An} {Automated} {Pipeline} for {Task} {Collection} and {Decontaminated} {Evaluation} of {Software} {Engineering} {Agents}},
	shorttitle = {{SWE}-rebench},
	url = {http://arxiv.org/abs/2505.20411},
	doi = {10.48550/arXiv.2505.20411},
	urldate = {2025-10-18},
	publisher = {arXiv},
	author = {Badertdinov, Ibragim and Golubev, Alexander and Nekrashevich, Maksim and Shevtsov, Anton and Karasik, Simon and Andriushchenko, Andrei and Trofimova, Maria and Litvintseva, Daria and Yangel, Boris},
	month = may,
	year = {2025},
	note = {arXiv:2505.20411 [cs]},
}

\end{document}